# Optimization of magneto-electric properties in Lead-free ($x$)$Co_{1.2}Ti_{0.2}Fe_{1.6}O_4$ - (100-$x$)$BaTiO_3$ based composites

Rajeev Dwivedi[1], Samanway Mohanta[2], Ashutosh Anand[3], Abhinash Tripathy[2], Najnin Bano[2], Dharmendra Kumar[4], Hemant Singh[2], R. Venkatesh[2], Sachin Gupta[1,*], Dinesh Kumar Shukla[2,#]

[1]Department of Physics, Bennett University, Greater Noida 201310, India
[2]UGC-DAE Consortium for Scientific Research, Indore 452001, India
[3]University School of Basic and Applied Science, Guru Gobind Singh Indraprastha University, Delhi 110078, India
[4]Department of Physics, Central University of Punjab, Bathinda, Punjab 151401, India

Email: *sachin.gupta@bennett.edu.in and #dkshukla@csr.res.in

**Abstract**

This work presents a systematic study of lead-free multiferroic composites of (x)$Co_{1.2}Ti_{0.2}Fe_{1.6}O_4$ - (100-x)$BaTiO_3$ (x = 10, 20, 30), which were synthesized by a solid-state reaction method to investigate the effects of composition and sintering temperature on their structural , electrical, magnetic, and magnetoelectric (ME) properties. X-ray diffraction along with Rietveld refinement confirms the coexistence of tetragonal $BaTiO_3$ (BTO) and cubic spinel $Co_{1.2}Ti_{0.2}Fe_{1.6}O_4$ (CTFO) phases. Microstructural analysis shows that densification and grain growth are better at higher sintering temperatures, leading to better coupling between the two phases. Dielectric and ferroelectric studies indicate lossy polarization-electric field (P-E) behaviour due to leakage from the conductive phase, while magnetic properties show increased magnetization with increasing ferrite content. All composites exhibit ME coefficients, which depend on the composition and sintering conditions; the highest ME coefficient (~1.28 mV/cm.Oe) was observed for the 30CTFO - 70BTO composite sintered at 1200 °C. This improvement is due to the optimal balance between magnetostrictive and piezoelectric responses and improved interfacial strain transfer. These results demonstrate that simultaneous optimization of dopant-modified composition and sintering conditions is essential for achieving improved magnetoelectric coupling in bulk multiferroic composites. Moreover, the results demonstrate the potential of lead-free composites for multifunctional device applications in next-generation, low-power technologies, including high-density non-volatile memory (e.g. FeRAM/MRAM), magnetic field sensors, spintronic devices, and actuators.

## Introduction

Multiferroic and magnetoelectric (ME) materials have attracted significant attention from scientists and engineers due to their intriguing physical properties and diverse applications [1-4]. In these materials, the coupling enables electric field control of magnetism and vice-versa. This makes them highly promising for next-generation multifunctional devices as well as fundamental studies [5,6]. Magnetoelectric materials can be synthesized as single-phase or composite materials. Single-phase materials typically exhibit weak ME coupling. In contrast, composite materials often produce strong ME responses due to interactions between different phases [5,7,8]. The ME coupling in composites largely depends on their microstructure and the quality of phase interfaces. However, challenges such as atomic diffusion and unwanted chemical reactions at phase interfaces can alter chemical and structural properties, which may result in reduction of coupling efficiency [1,9]. The synthesis of ME composite materials involves combining components that individually possess these properties, ferroelectricity and ferri/ferromagnetism. In these composites, the magnetostrictive phase and the piezoelectric phase interact, enabling the control of electric polarization using a magnetic field or the control of magnetization using an electric field [8,10-12]. These effects are known as the direct and indirect ME effects and hold significant potential for various applications, including spintronics devices, transducers, sensors, medical devices and magnetoelectric random access memories (MERAMs) [13-17]. Achieving a strong ME effect requires high magnetostriction, excellent piezoelectricity, and efficient interfacial coupling between phases. Among the various systems, $CoFe_2O_4$-$BaTiO_3$ (CFO-BTO) composites are widely studied because they have complementary properties [18,19]. $CoFe_2O_4$ (CFO) is a ferrimagnetic ceramic with a partially inverted spinel structure ($AB_2O_4$), It exhibits a high Curie temperature, strong magnetocrystalline anisotropy, high coercivity, high saturation magnetization, and a significant magnetostrictive coefficient [19-22]. It is chemically stable, mechanically strong, and exhibits hard magnetic behavior. In CFO, $Co^{2+}$ ions are located mainly at octahedral (B) sites, while $Fe^{3+}$ ions are distributed between tetrahedral (A) and octahedral (B) sites in a dense-packed oxide lattice [23,24]. Whereas $BaTiO_3$ (BTO), a perovskite compound ($ABO_3$), exhibits excellent dielectric, ferroelectric and piezoelectric properties [25-28]. Its crystal structure allows cation substitution without significant structural distortion. BTO undergoes phase transitions from cubic to tetragonal, orthorhombic, and rhombohedral phases, with the tetragonal phase being stable below ~120 °C and the cubic phase being stable above this temperature [25,29-32].

Many recent studies have focused on enhancing magnetoelectric coupling in various composite systems, such as PVDF-TrFE/$CoFe_2O_4$, PVDF-$ZnFe_2O_4$, Ni-$BaTiO_3$, Co/Fe-$Ba_{0.7}Sr_{0.3}TiO_3$, $BiFeO_3$-$CoFe_2O_4$, $CoFe_2O_4$/PZT, and $CoFe_2O_4$/$BaTiO_3$, through chemical doping and structural modification [14,23,33-37]. These studies show that ME coupling is highly dependent on microstructure, phase distribution, and interface quality. However, achieving high ME coupling in bulk CFO-BTO composites remains a challenge due to interfacial diffusion, formation of secondary phases, and inefficient strain transfer during processing at high temperatures. Although doping and compositional tuning can change structural and magnetic properties, they often have disadvantages such as reduced ferroelectric performance or increased leakage current. Moreover, laminate-based systems have shown significantly higher ME coefficients, however, their complex synthesis limits their practical applications compared to bulk ceramics. Recent advances in multiferroic materials emphasize the importance of interface engineering, strain-mediated coupling, and controlled processing conditions to enhance the ME response. Nevertheless, systematic studies that simultaneously consider dopant modification, compositional variation and sintering optimization in CFO-BTO bulk composites are limited. Therefore, there is a clear need to develop strategies that improve both individual phase properties and interfacial coupling efficiency. Ti doping in CFO is particularly important due to its ionic radius being comparable to that of $Fe^{3+}$, enabling substitution with minimal structural disruption [38]. Ti substitution significantly influences the magnetic and structural properties of CFO and can enhance its performance in ME composites [39-41].

Thus, in the present work, Ti-doped CFO and BTO are selected as the magnetostrictive and piezoelectric phases, respectively, to investigate the combined effects of composition and sintering temperature. In this study, the structural, morphological, electrical, magnetic and magnetoelectric properties of (x)$Co_{1.2}Ti_{0.2}Fe_{1.6}O_4$ - (100-x)$BaTiO_3$ composites (x = 10, 20, 30) have been systematically investigated. The effect of composition and sintering temperature on these properties has been investigated in detail, with a focus on enhancing magnetoelectric coupling. The novelty of this work lies in the simultaneous optimization of the dopant-modified composition and processing conditions. Furthermore, a clear correlation between microstructure, phase evolution, and magnetoelectric response has been established, providing insights for the design of high-performance multiferroic materials.

## 1. Experimental details

Magnetoelectric composites constituting Ti-substituted $CoFe_2O_4$ as a ferromagnetic and $BaTiO_3$ as ferroelectric phase were synthesized via solid state reaction method. $Co_3O_4$, $Fe_2O_3$, and $TiO_2$ (these are ~99.9% purity) powders were precisely weighed in their stoichiometry ratios and thoroughly mixed using an agate mortar and pestle for several hours, to ensure homogeneity and uniform elemental distribution. The homogenized mixed powder was calcined at 800 °C. After calcination, the powder was regrind to obtain a homogeneous fine powder of the desired compound. Commercially available $BaTiO_3$ (~99.95% purity) powder was used for synthesis of composites. For composites having specific formula $(x)Co_{1.2}Ti_{0.2}Fe_{1.6}O_4$ - $(100\text{-}x)BaTiO_3$ where x = 10, 20, 30 are prepared by mixing of CTFO and BTO. Hereafter, these samples are designated as 10CTFO-90BTO (C10:B90), 20CTFO-80BTO (C20:B80), and 30CTFO-70BTO (C30:B70). CTFO and BTO powders mixed with polyvinyl alcohol (PVA) which acts as a binder to hold the powders together during pressing and 10 mm diameter pellets were made using a hydraulic pelletizer, applying a force of 100 kilo Newton and these pellets are sintered at 1100°C and 1200 °C for 24 hours with controlled heating and cooling rate of 5 °C/min. Crystal structure and phase formation of all pellets are analysed using Bruker X-ray diffractometer utilizing Cu k-alpha radiation. Microstructural of all pellets are analysed using field emission scanning electron microscope (FESEM). A 16 Tesla vibrating sample magnetometer (VSM) of Quantum Design has been used for the magnetization measurements. For electrical measurements, these pellets were polished after which high purity silver paste has been used for electrodes. Dielectric measurements were carried out in the frequency range of 4 Hz to 1 MHz and the temperature range of 100-500 K using HIOKI IM 3536 LCR meter. Poling was performed at room temperature in silicon oil by applying a DC electrical field in range of 1-1.5 kV/mm for 30 minutes, where applied voltage was adjusted according to sample thickness to avoid dielectric breakdown. P-E hysteresis loops of the composites were measured using home developed setup [42]. The magneto-electric (ME) response of all poled samples have been characterized using lock-in amplifier-based technique, where small ac magnetic field (superimposed on dc magnetic field) was produced using a Helmholtz coil while dc magnetic field is produced using an electromagnet.

# 2. Results and discussions

## 3.1 Structural properties

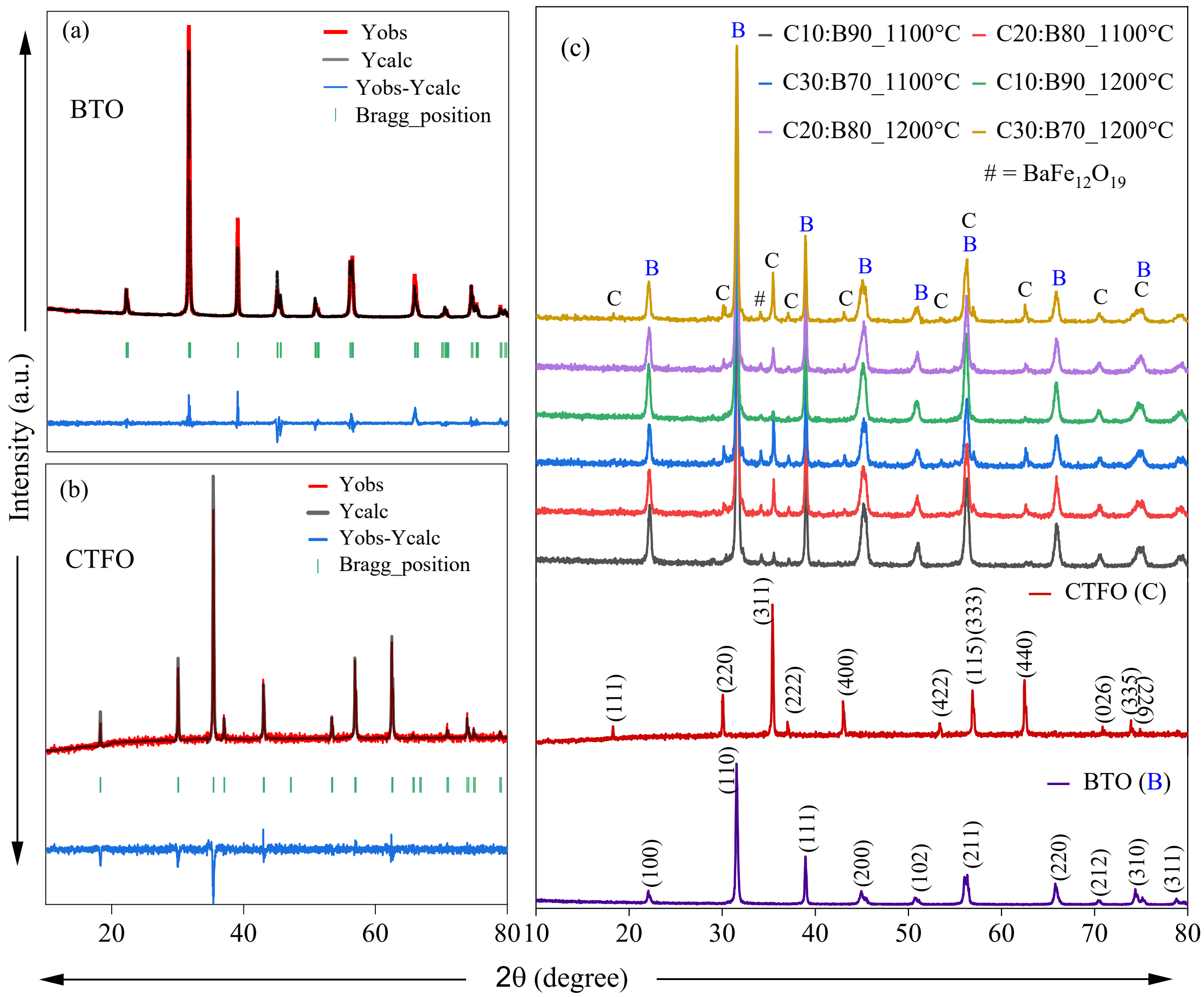

Fig. 1. Rietveld-refined X-ray diffraction patterns for (a) pure BTO, and (b) pure CTFO. (c) XRD patterns of pristine CTFO-BTO and composites [(x) $Co_{1.2}Ti_{0.2}Fe_{1.6}O_4$ - (100-x) $BaTiO_3$ x= 10, 20, 30] at room temperature.

Fig. 1 shows the room-temperature X-ray diffraction (XRD) patterns of all [(x) $Co_{1.2}Ti_{0.2}Fe_{1.6}O_4$ -(100-x) $BaTiO_3$ x= 10, 20, 30] samples along with the Rietveld refined patterns of the individual phases of CTFO and BTO. The XRD patterns of pure CTFO and BTO reveal crystallization in the cubic inverse spinel structure and tetragonal perovskite structure, respectively. These results are consistent with previous reports [38,43-45]. In all composite, reflections corresponding to both the ferroelectric (BTO) and ferromagnetic (CTFO) phases are clearly visible in the XRD patterns, confirming the successful formation of a two-phase composite system**.** Beyond the primary phases, the XRD patterns reveal a low-intensity reflection (marked as #), indicating a subtle interfacial reaction between CTFO and BTO during high-temperature sintering. This secondary phase is a well-documented characteristic of the CTFO-BTO system [46-48]. Given its reported magnetic nature, we propose that this interfacial phase may facilitate the strain-mediated coupling, thereby contributing positively to the overall magneto-electric response of the composites. Furthermore, the intensity of the ferromagnetic (spinel) phase peaks increases systematically with increasing CTFO content, which is clearly evident in Fig. 1(c).

Fig. 2. XRD patterns along with Rietveld refinement for CTFO-BTO composites with compositions (a,d) C10:B90, (b,e) C20:B80, and (c,f) C30:B70 sintered at 1100 °C (a–c) and 1200 °C (d–f).

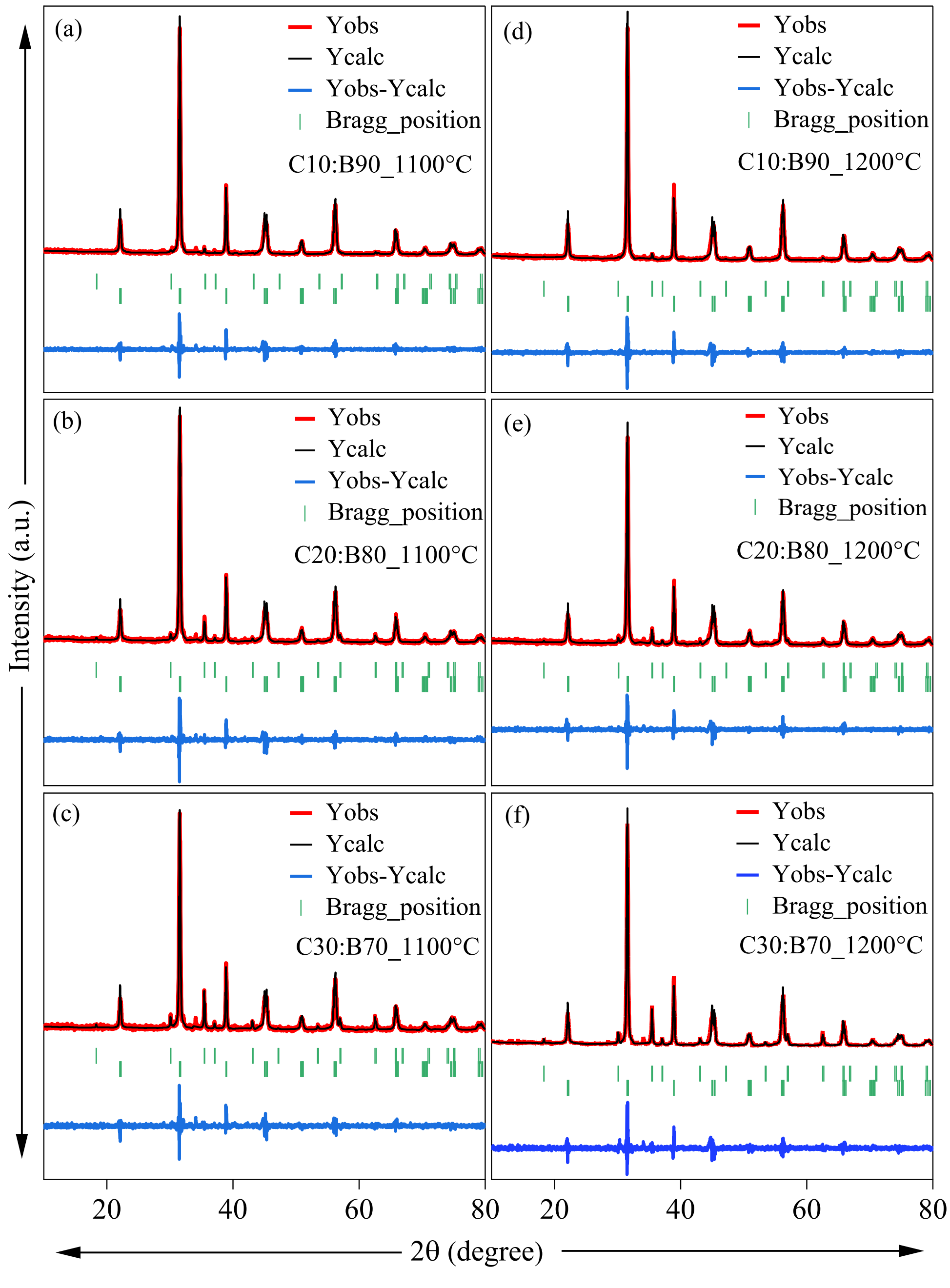

Rietveld refinement of the XRD data of composite samples were also carried out using a two-phase model using FullProf software. Fig. 2 (a-f) presents Rietveld refinements of CTFO-BTO composite samples. The refinement was performed using the P4mm space group for the ferroelectric BTO phase and Fd-3m for the ferrite CTFO phase. The refined lattice parameters and Rietveld reliability factors are summarized in Table S1. Pure BTO exhibits lattice parameters of a = b = 3.993 Å and c = 4.035 Å, yielding a c/a ratio of 1.01, characteristic of its tetragonal symmetry. The CTFO phase shows a cubic lattice with a = 8.401 Å. In the composite samples, the BTO phase retains its tetragonal structure with c/a values in the range of approximately 1.007-1.008, while the CTFO lattice parameter slightly varies between 8.365 and 8.397 Å. The $\chi^2$ values for all composites lie close to unity (≈ 1.6-1.9), confirming the reliability and robustness of the refinements. Overall, the coexistence of the perovskite BTO and spinel CTFO phases in all composites sintered at 1100 °C and 1200 °C demonstrates good phase stability and chemical compatibility between the two components.

### 3.2 Morphological studies

Fig. 3 shows the FESEM micrographs and corresponding histograms of CTFO-BTO composites sintered at two different temperatures. The first set of Fig. 3 (a, b, c, g, h and i) corresponds to samples sintered at 1100 °C, while the second set of Fig. 3 (d, e, f, j, k and l) depicts the microstructure of samples sintered at 1200 °C. These FESEM images were analyzed using Image-J software. All images show uniformly distributed micron-sized crystallites, confirming the formation of a homogeneous composite phase. As shown in Fig. 3, the grain size distribution of all composite samples exhibits significant grain growth with increasing sintering temperature. The average grain size is approximately ~1.0 μm for the composite sintered at 1100 °C and approximately ~1.3 μm for the composite sintered at 1200 °C, which can be attributed to increased atomic diffusion and grain coalescence. Higher sintering temperatures provide sufficient activation energy, facilitating grain boundary migration and enabling grains with lower surface energy to grow at the expense of smaller grains [49]. Consequently, the composite sintered at 1200 °C possesses a denser microstructure with lower porosity, which is expected to contribute to improved piezoelectric and mechanical properties. The densification behaviour i.e. bulk density of sintered pellets, relative density, and porosity were calculated by geometrical method. The values are shown in Table S2. The bulk density of all these composites was found to be in the range of ~4.5-5.2 g/cm³, corresponding to a relative density of ~77-89%, depending on thickness and sintering conditions. Composites sintered at 1200 °C exhibit improved densification and reduced porosity, contributing to enhanced coupling between both phases and improved magnetoelectric response.

Fig. 3. FE-SEM micrographs of CTFO–BTO composites: C10:B90, C20:B80, and C30:B70 sintered at 1100 °C (a–c) and 1200 °C (d–f), along with corresponding grain size distribution histograms (g–l).

### 3.3 Dielectric properties

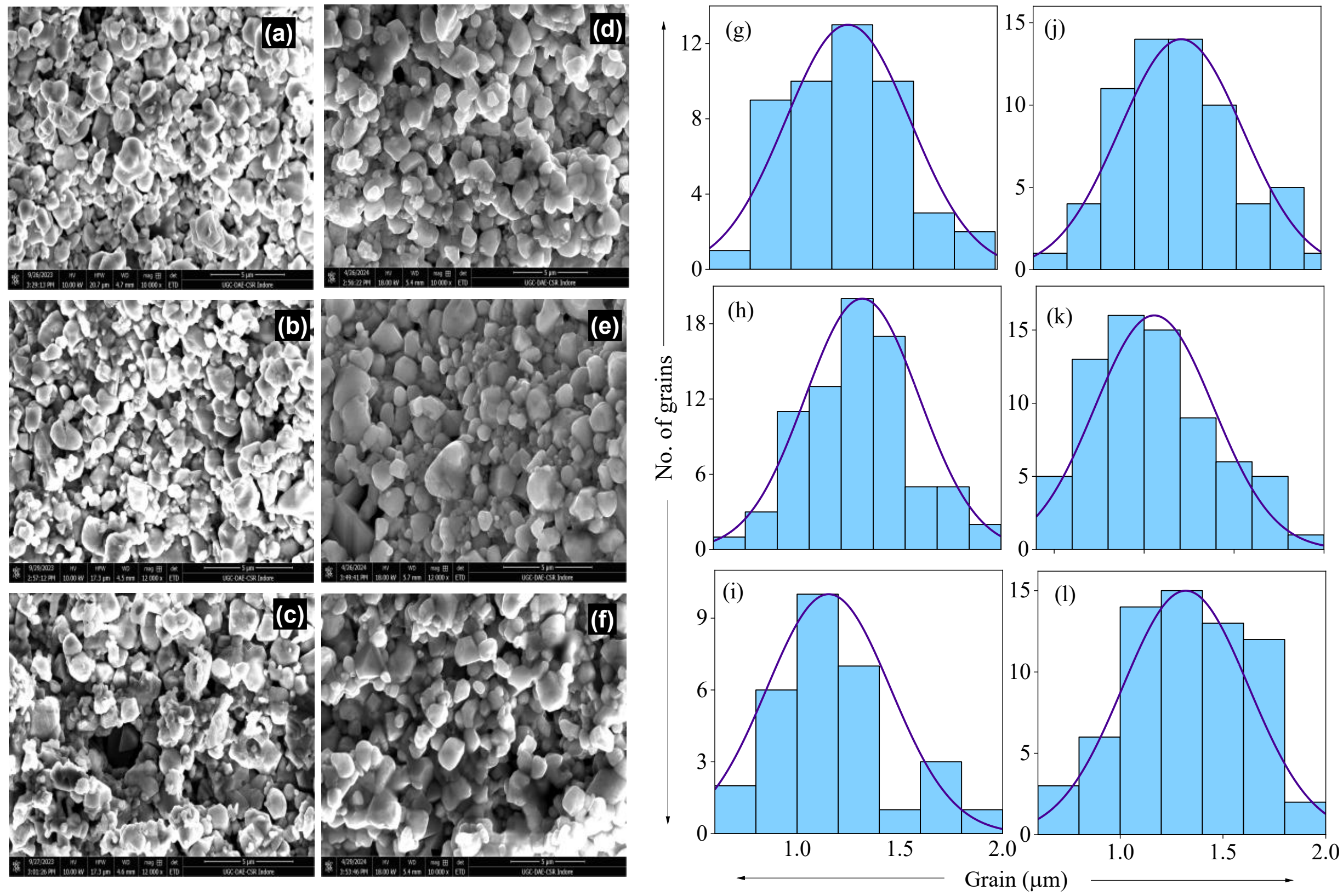

Dielectric measurements were conducted using the parallel capacitor principle. The dielectric properties of all composites were studied by evaluating the frequency dependent capacitance (C) and dielectric loss (tanδ) at different temperatures. Subsequently, capacitance and tanδ were examined as functions of temperature across various frequencies. All composites were measured over a wide frequency range from 4 Hz to 1 MHz, with temperature variations from 100 K to 500 K. The real dielectric constant (ε') of all composites was calculated using the following equation (i):

$$\varepsilon' = \frac{C.t}{\varepsilon_o.A} \quad \text{(i)}$$

Here C represents the capacitance, t is the sample thickness, A is the cross-sectional area, and $\varepsilon_o$ is the permittivity of free space ($8.854 \times 10^{-12}$ F/m) [49-52].

### 3.3.1 Temperature dependent dielectric properties

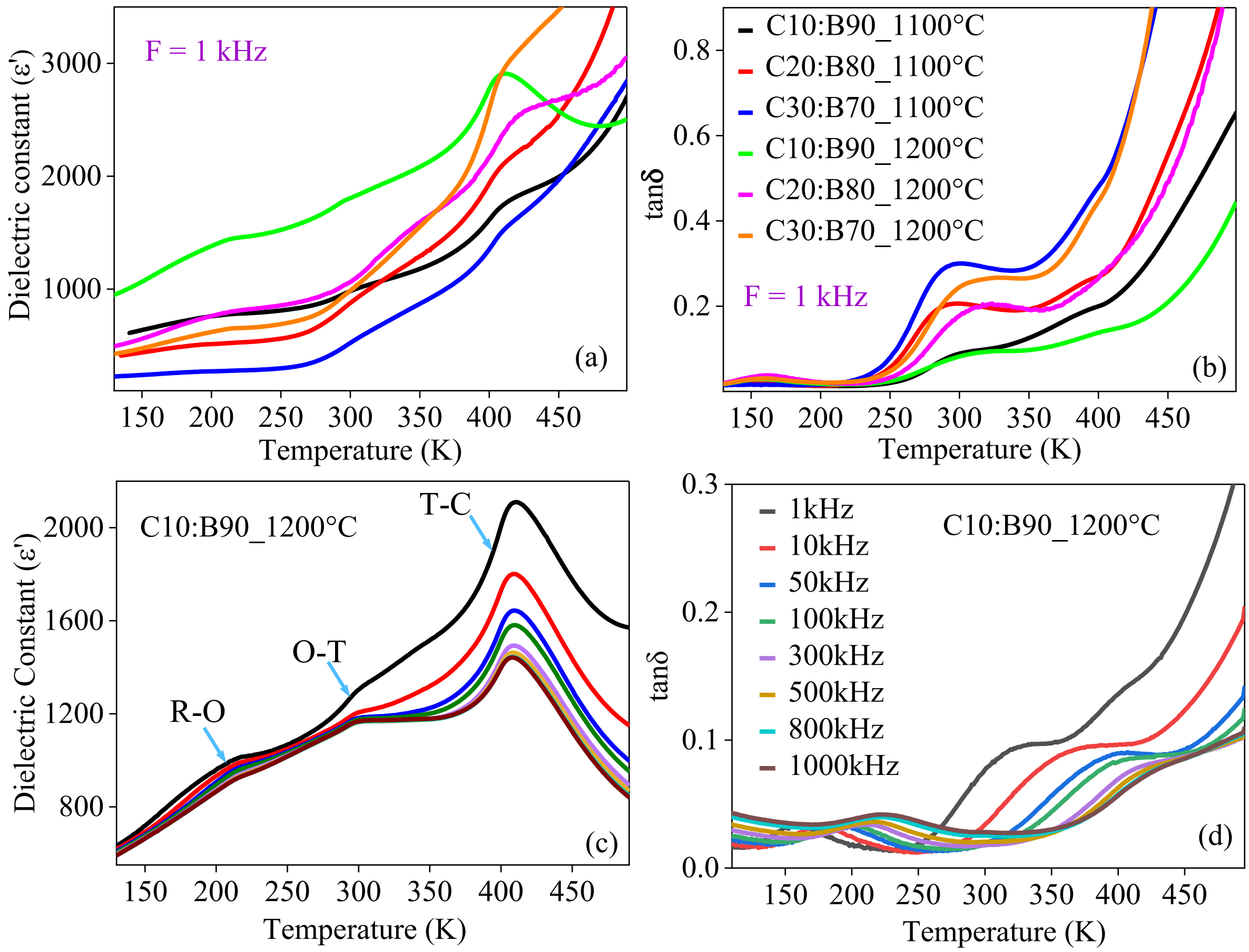

Fig. 4. Temperature dependence of (a) dielectric constant (ε') and (b) dielectric loss (tanδ) for CTFO-BTO composites with 10%, 20%, and 30% wt.% of CTFO at two different sintering temperatures, at a constant frequency of 1 kHz. Temperature dependence of (c) dielectric constant (ε') and (d) dielectric loss (tanδ) for composite C10:B90_1200°C as a function of frequency.

The temperature dependence of the dielectric constant (ε′) and dielectric loss (tanδ) of the composites measured at a constant frequency of 1 kHz is shown in Fig. 4 (a and b), respectively. Fig. 4 (c and d) present the function of frequency dielectric constant (ε′) and dielectric loss (tanδ) of the C10:B90 composite sintered at 1200 °C. Below the ferroelectric-paraelectric transition temperature (~400 K), the dielectric constant decreases with increasing CTFO content (10-30 wt%), reflecting the intrinsically lower permittivity of CTFO compared to ferroelectric $BaTiO_3$ (BTO). For a fixed composition, samples sintered at 1200 °C exhibit approximately 25% higher ε′ than those sintered at 1100 °C. This enhancement is attributed to improved densification, which strengthens Maxwell-Wagner interfacial polarization between insulating BTO grains and comparatively more conductive CTFO grains. At certain temperatures, materials undergo phase transitions from ferroelectric to paraelectric, which cause significant changes in the dielectric constant (ε') [53]. As the temperature increases, the dielectric constant increases, attributed to thermally activated polarization, improved dipole alignment, and domain wall contributions, and reaches a maximum at Curie temperature ($T_c$). Above $T_c$, the dielectric constant decreases due to a phase transition from the ferroelectric to the paraelectric, where the ferroelectric domains disappear. In Fig. 4 (a and c), several transitions are observed as the temperature changes from 100 K to 500 K. The phase transitions between rhombohedral, orthorhombic, tetragonal and cubic occurring at 210, 300 and 400 K can be clearly seen in the Fig. 4(c). In Fig. 4(b), the temperature-dependent dielectric loss shows an increasing trend with a higher content of CTFO. Additionally, the dielectric loss reduces with higher sintering temperatures. The variation with temperature reveals that at higher temperatures, the dielectric loss becomes significantly larger. The relatively low dielectric loss at lower temperatures can be attributed to strong ionic bonding, rigid crystal structures that suppress dipole relaxation, reduced defect concentration, and limited interfacial polarization in the composite system [44,54-56]. Higher sintering temperature reduces porosity and improve crystalline quality leading to low dielectric loss.

As shown in Fig. 4(c), the dielectric constant exhibits higher values at low frequencies and decreases with increasing frequency. This behaviour arises from the inability of space-charge and interfacial polarization mechanisms to follow the alternating electric field at higher frequencies. The C10:B90 composite sintered at 1200 °C shows a clear frequency-dependent dielectric response, with the temperature corresponding to the dielectric maximum shifting toward higher temperatures with increasing frequency. This shift indicates relaxer-like ferroelectric behavior with diffuse phase transitions. Fig. 4 (d) demonstrates that dielectric loss decreases with increasing frequency, as charge hopping and interfacial polarization processes cannot respond effectively at higher frequencies. The monotonic increase of tanδ with temperature, particularly above $T_c$, is attributed to thermally activated charge carriers. The frequency-dependent shift in transition temperature further confirms the relaxer ferroelectric nature of the composite.

### 3.3.2 Frequency-dependent dielectric properties

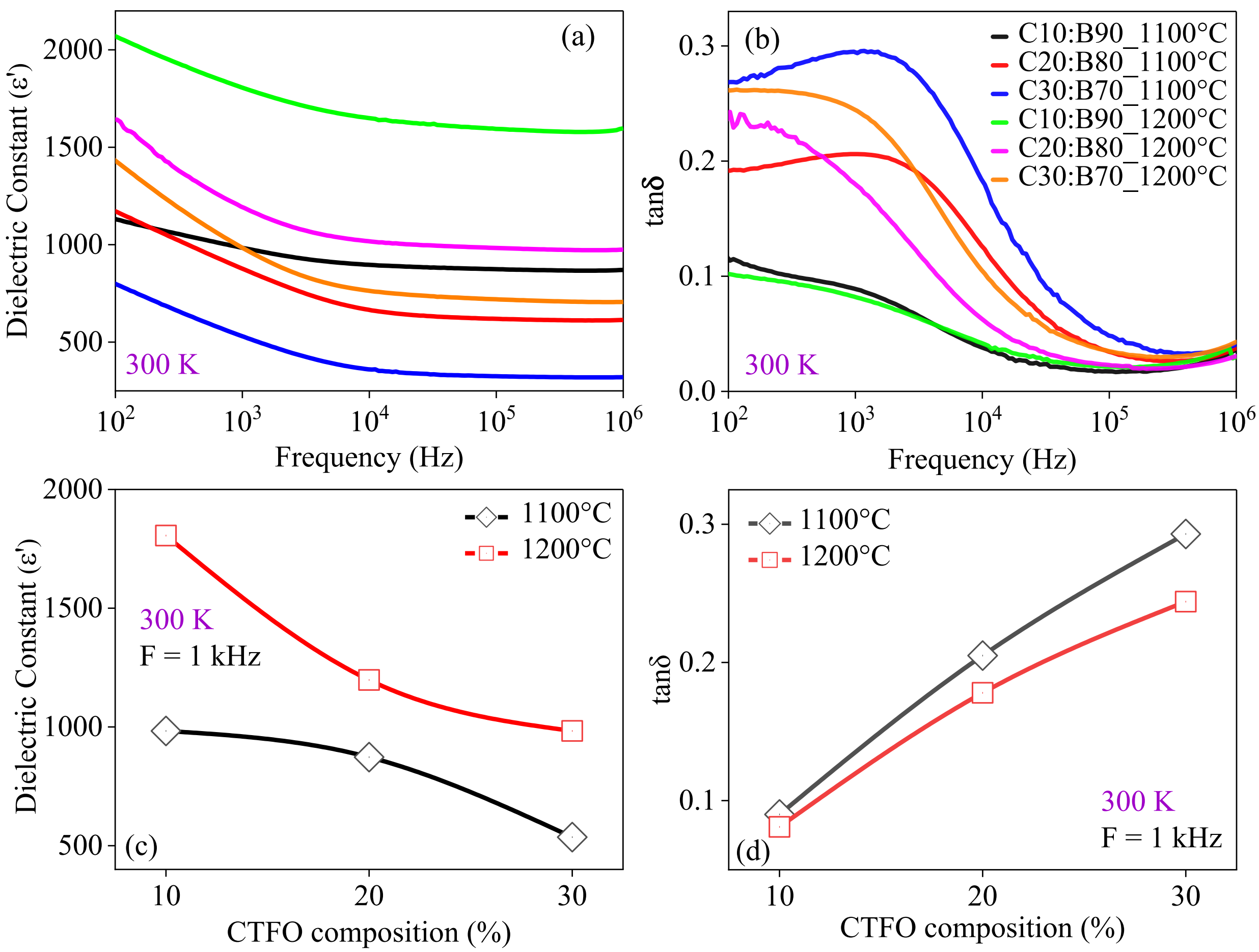

Fig. 5. Frequency dependence of (a) Dielectric constant (ε'), (b) dielectric loss (tanδ) for CTFO-BTO composites with 10%, 20%, and 30% of CTFO at 300 K, and (c) and (d) illustrate the variation of ε' and tanδ as a function of composition with sintering temperature.

Fig. 5 and Fig. S2 shows the frequency dependence ($10^2$–$10^6$ Hz) of the dielectric constant (ε′) and dielectric loss (tanδ) of the composites measured at room temperature. At lower frequencies, the dielectric constant exhibits high values with strong low-frequency dispersion for all composites. With increasing frequency, ε′ gradually decreases and eventually becomes nearly constant at higher frequencies. This behavior can be explained using the Maxwell-Wagner interfacial polarization model [56,57]. At low frequencies, interfacial polarization, space-charge polarization, and grain-boundary effects contribute significantly to the dielectric response. At higher frequencies only intrinsic lattice polarization remains active, as charge carriers and dipoles are unable to follow the rapid reversal of the applied electric field. The dielectric constant of the composites lies between those of pure BTO and pure CTFO, systematically decreasing with increasing ferromagnetic phase content due to dilution of the ferroelectric BTO phase which is clearly observable in Fig. 5 (c). Pure BTO exhibits the highest dielectric constant among all samples, as summarized in Table S3. Furthermore, Fig. 5 (c and d) clearly shows increasing the sintering temperature leads to an enhancement in $\varepsilon'$, which is attributed to improved grain connectivity, reduced porosity, and stronger interfacial polarization effects. Fig. 5 (b) indicates dielectric loss reduces with increasing frequency due to a higher frequencies, charge carriers do not follow rapidly alternating electric field, and at lower frequencies, dielectric loss exhibits a maximum because enhanced charge carrier hopping and interfacial polarization then gradually decrease with increasing frequency [52,58]. The higher ferroelectric content composite shows low dielectric loss, which is attributed to the higher resistivity of the ferroelectric phase (BTO) compared to CTFO. Very interesting feature observed in these composited is increasing the sintering temperature decreases the $\tan\delta$ as is clear seen in Fig. 5 (d). Since higher sintering temperature improves the microstructural quality of the samples by reducing porosity, promoting grain growth and less energy loss as heat and reduces leakage paths at grain boundaries. This type of nature is very useful for high frequency applications such as in the development of devices like RF, microwave circuits etc.

### 3.3.3 Impedance analysis

Impedance analysis is a powerful technique used to investigate the electrical behavior and conduction mechanisms of materials. The real ($Z'$) and imaginary ($Z''$) components of impedance are calculated using the following equation (ii):

$$Z' = \frac{\varepsilon''}{\omega C_0\left(\varepsilon'^2+\varepsilon''^2\right)} \quad \text{And} \quad Z'' = \frac{\varepsilon'}{\omega C_0\left(\varepsilon'^2+\varepsilon''^2\right)} \qquad \text{(ii)}$$

where, $C_0$ is the vacuum capacitance, ω is the angular frequency, ε′ and ε″ are the real and imaginary parts of the dielectric constant, respectively [52].

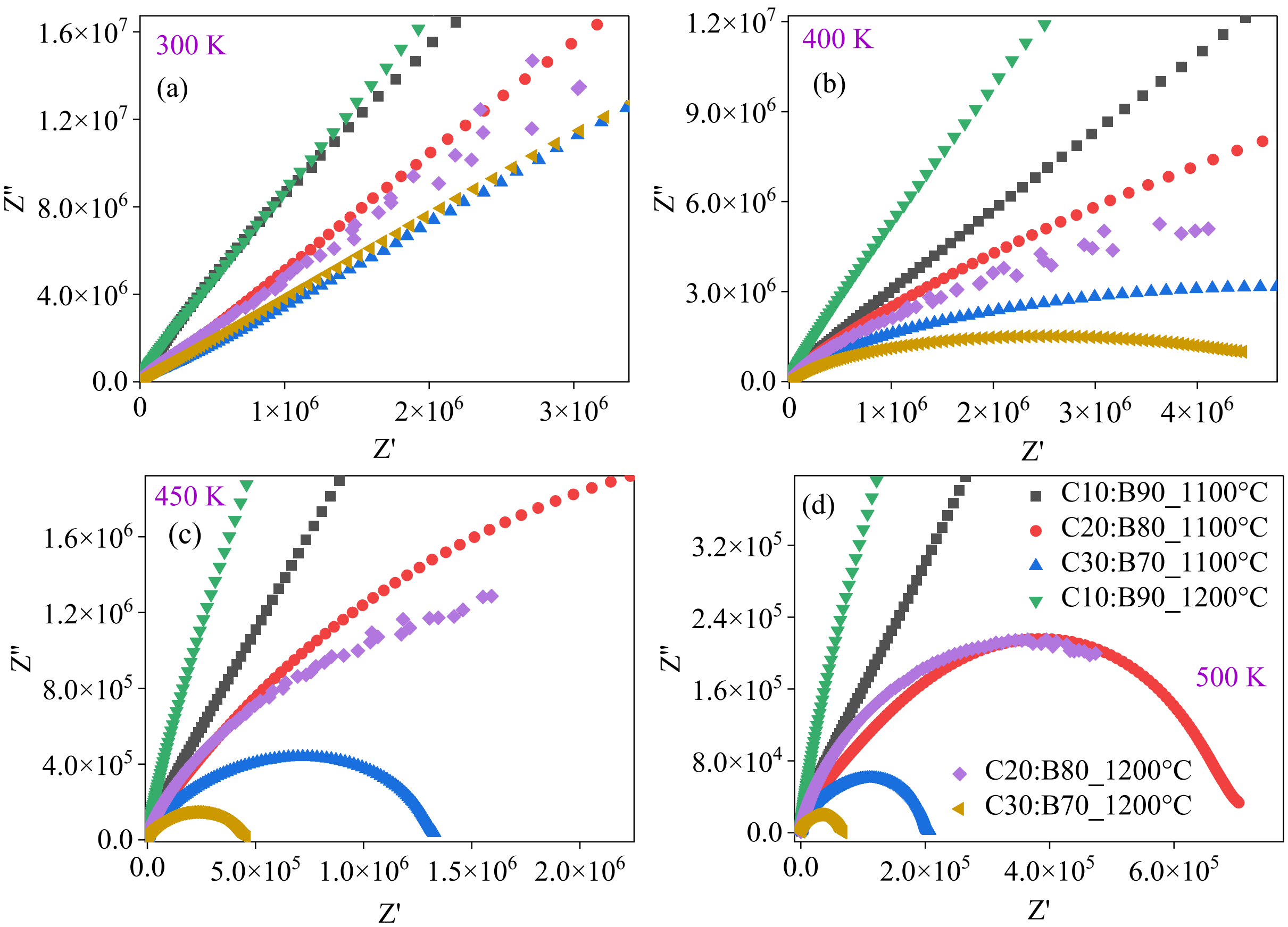

Fig. 6. Nyquist plots (Z″ vs Z′) of CTFO–BTO composites sintered at 1100 °C and 1200 °C, measured at selected temperatures (a) 300 K, (b) 400 K, (c) 450 K, and (d) 500 K.

The Nyquist plots of (Z′ and Z″) of composite (x)$Co_{1.2}Ti_{0.2}Fe_{1.6}O_4$ - (100-x) $BaTiO_3$ x= 10, 20, 30 and sintered at 1100 °C and 1200 °C a wide range of frequency 5 Hz-1 MHz at four different temperatures 300 K, 400 K, 450 K, and 500 K are show in Fig. 6 (a-d). At room temperature (300 K), the Nyquist study exhibits highly inclined straight-line behaviour without the formation of a semicircle, indicating a highly resistive nature of all composites. This type of behaviour is mainly attributed to the dominance of electric polarisation and the insulating characteristics of the composite. In Fig. 6 (b and c), 400 K to 450 K impedance spectra start to show curvature and at 500 K, depressed semicircular arcs are clearly observed in Fig. 6 (d). These arcs represent the presence of bulk relaxation processes such as grains, grain boundaries, and distinct crystallographic phases in the composites [59]. Linear to semicircular transition indicates activation of conduction mechanisms with increasing temperature (400 K - 500 K). Further, in Fig.

6 it is observed that increasing the concentration of the ferrite phase (CTFO) leads to a decrease in resistance due to the introduction of conductive ferrite. The Composite sintered at 1200 °C exhibits noticeable variation in resistance, indicating the influence of sintering temperature on microstructure, densification, and grain boundary characteristics. At 300 K, the Nyquist plot shown in Fig. S3 was analyzed using an equivalent circuit model. The impedance data (Z′ and Z″) were fitted using EIS Spectrum Analyzer software with a model consisting of a constant phase element (CPE) in parallel with a resistor (R). CPE is used to indicate deviation of capacitance from ideal behaviour. The values of R and CPE obtained from fitting are summarised in Table S4. It can be observed that the experimental data and fitted data show that the composite was highly insulating ($10^8$–$10^9$ Ω) at room temperature, clearly indicating resistance dependence on composition and sintering temperature. Moreover, the non-Debye relaxation behaviour present in composite materials is justified by the presence of the CPE exponent (n), which lies between 0.85 and 0.93, attributed to structural heterogeneity and grain boundary effects. The CPE parameter (P) range of $10^{-10}$ F is consistent with dielectric relaxation processes in composites. Increasing the CTFO content in the composite reduces the resistance and enhances the magnetic properties. They provide an optimal balance between the two phases to achieve maximum magnetoelectric coupling.

### 3.4 Ferroelectric properties

The ferroelectric behaviour of CTFO-BTO composite ceramics was analysed at room temperature. The polarization versus electric field (P-E) hysteresis loops for all composite samples and pure BTO were recorded under a maximum electric applied electric field, adjusted for each sample to obtain saturation while avoiding dielectric breakdown at a low frequency of ~50 Hz, as illustrated in Fig. 7. The P-E loop of all composites exhibits unsaturation and lossy behaviour observed, which indicate leakage current. The presence of hysteresis loop confirms the ferroelectric nature of all composite materials. The saturation polarization ($P_s$) decreases with increasing CTFO content due to the non-ferroelectric ferrite. Additionally, a change in sintering temperature enhances the saturation polarization of each sample by approximately 30%, as clearly shown in Fig. 7 (c). The coercive field (Ec) increases with increasing CTFO content, due to increased domain wall pinning and internal stress from the ferromagnetic phase, while the higher conductivity of CTFO also contributes to the leakage effect. However, for some composites, an increasing sintering temperature leads to a decrease in $E_c$, because higher temperatures reduce defects and promote grain growth, resulting in fewer grain boundaries, as clearly shown in Fig. 3 While Fig. 7(d) clearly shows remanent polarization change in variation of CTFO, increasing the sintering temperature $P_r$ also changes, $P_r$ remains above 0.76 mC/cm$^2$ across all composites confirming stable ferroelectric domain switching.

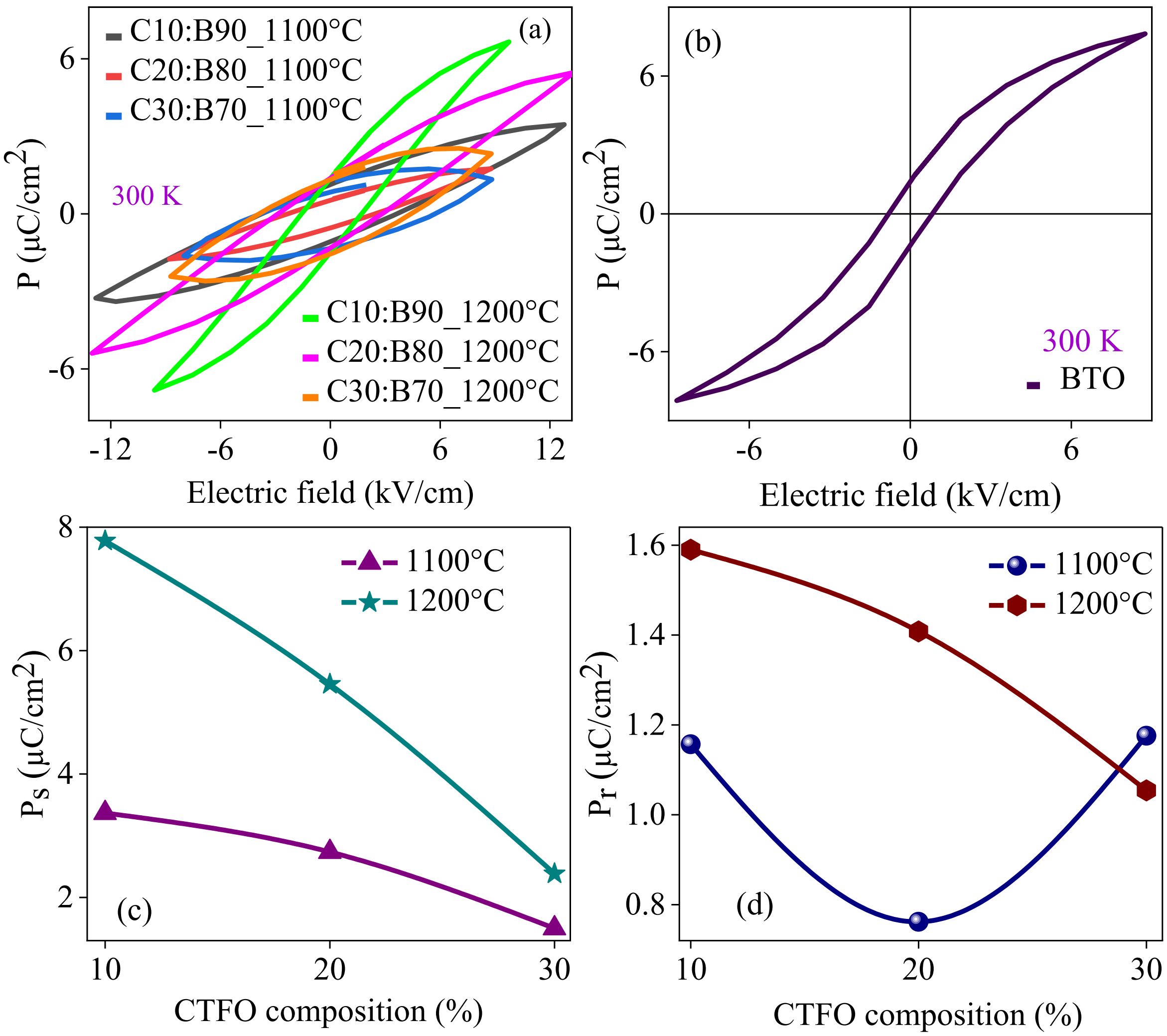

Fig. 7. The electric field dependence of polarization for (a) (x)CTFO – (100-x)BTO (b) for BTO, measured at room temperature. Composition dependence of (c) saturation polarization and (d) remnant polarization as a function of sintering temperature.

Pure BTO exhibits a lower coercive, high saturation polarization and remanent polarization field compared to the CTFO-BTO composites, as evident in Fig. 7 (b). In our composite higher sintering temperature exhibit low $E_c$, high $P_s$ and $P_r$. This improvement of all composites is attributed to enhanced densification and improved crystallinity at higher sintering temperatures, which are beneficial for the ferroelectric properties of ceramic. Pure BTO shows perfectly saturated ferroelectric behavior, whereas all composite samples exhibit lossy hysteresis loop shapes. This behaviour arises because the ferroelectric properties are progressively diluted by the incorporation of the ferromagnetic phase. Overall, the results demonstrate significant variations in $P_s$, $P_r$ and $E_c$ with increasing CTFO content and sintering temperature, as listed in Table 1. Ferroelectric properties are influenced by composition, homogeneity, defects, external fields, and sintering temperature, all of which contribute to the material's response, as clearly shown in Fig. 7 (c and d).

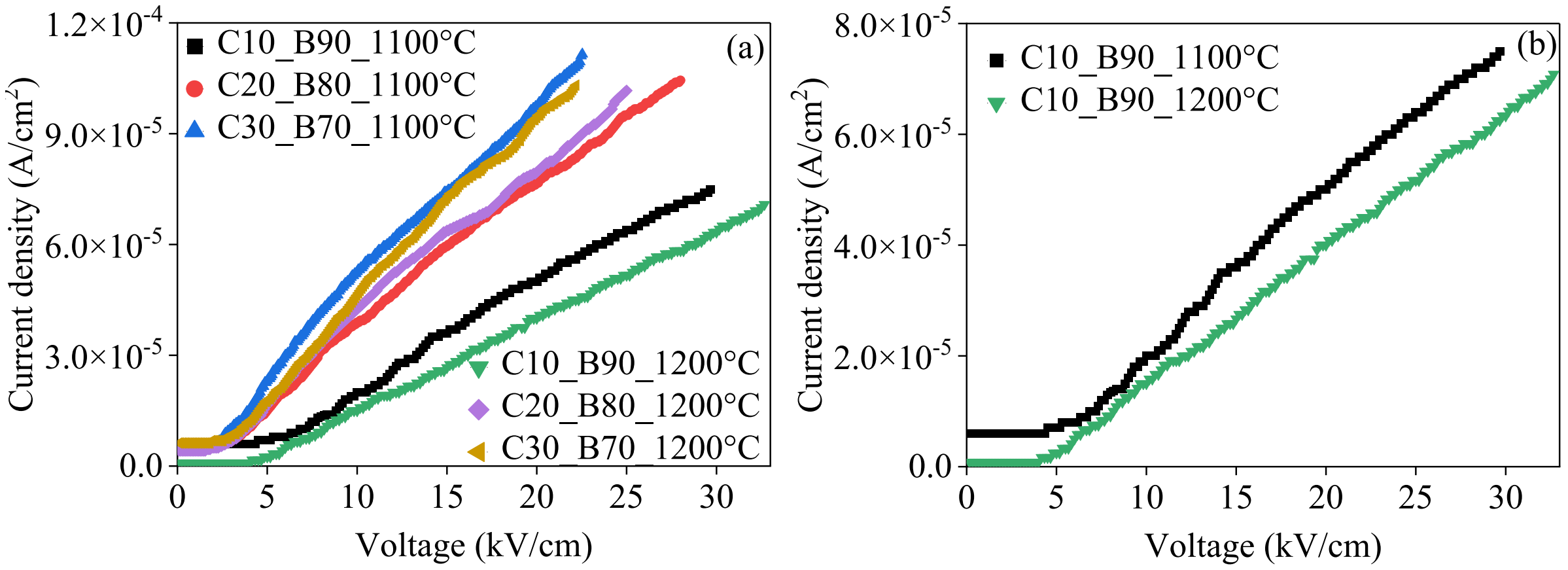

Fig. 8. (a) Current density–electric field (J-E) behaviour of CTFO-BTO composites (x = 10, 20, 30) at 1100 °C and 1200 °C (b) Comparison of J-E characteristics for C10:B90 at different sintering temperatures.

Fig. 8 illustrates the leakage current characteristics of CTFO-BTO composites at different sintering temperatures (1100 °C and 1200 °C) and varying compositions as a function of applied electric field. The leakage current increases with increasing CTFO content, with the highest current density observed in the C30:B70 composite. This is because the resistivity of ferromagnetic materials is generally lower than that of ferroelectric materials. For all compositions, composites sintered at 1200 °C exhibit lower current density than those sintered at 1100 °C, which can be attributed to better densification, lower porosity, and improved grain boundary characteristics. Fig. 8(b) demonstrates the effect of sintering temperature, showing reduced leakage current at higher temperatures. This behavior indicates that, depending on composition, a higher sintering temperature improves insulation properties. Lower leakage current at higher sintering temperatures enables better ferroelectric response and improved magnetoelectric coupling.

### 3.5 Magnetic properties

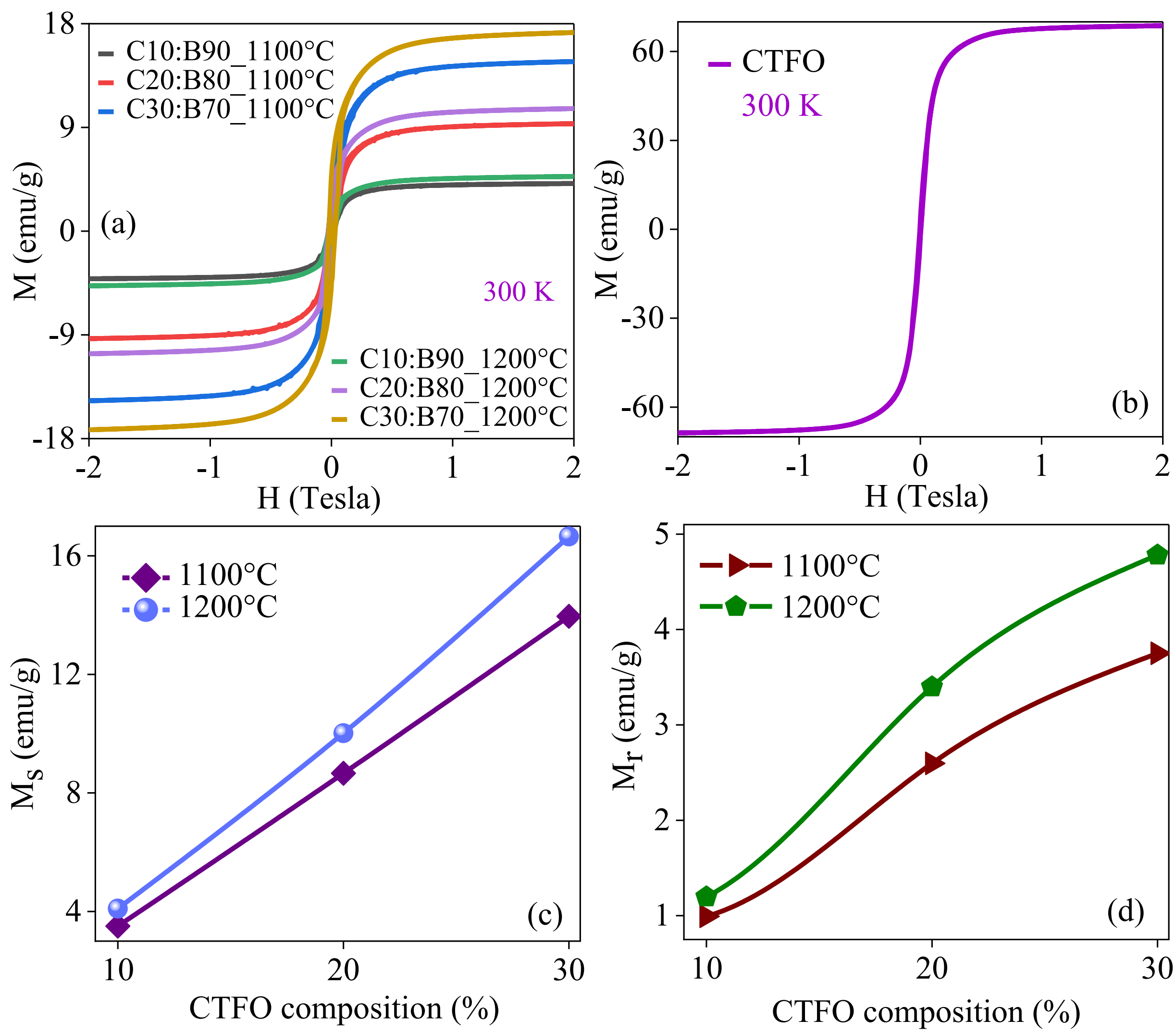

Fig. 9. (a) The magnetic field (H) dependence of magnetization (M) for (a) different composites and for (b) CTFO, measured at room temperature Composition dependence of (c) saturation magnetization ($M_s$) and (d) remanent magnetization as a function of sintering temperatures.

The magnetic measurements for composites CTFO-BTO (with $x$ = 10, 20 and 30) were analyzed at room temperature under an applied magnetic field of ± 2 T. The magnetic hysteresis loops for these composites sintered at 1100 ºC and 1200 ºC with varying CTFO content are depicted in Fig. 9 (a). Fig. 9 (c and d) show the dependence of saturation magnetization ($M_s$) and remnant magnetization ($M_r$) on the ferrite content. All composites show clear hysteresis loop, indicating ferromagnetic behavior. Fig. 9 (b) shows the field dependence of magnetization for CTFO, the saturation magnetization ($M_s$), magnetic retentivity ($M_r$) and coercive field ($H_c$) values were found to be 67.5, 6.5 emu/g and 87.8 Oe, respectively. These magnetic observations align with previous reports, the findings of Praveen et al. [38], who reported a decrease in magnetization with $Ti^{4+}$ substitution and improved strain sensitivity. Due to the occupation of non-magnetic $Ti^{4+}$ ions at octahedral (B) sites, which weakens the A-B superexchange interaction. In the spinel structure, the net magnetization results from the difference between the magnetic moments of the $Fe^{3+}$ ions at tetrahedral (A) and $Co^{2+}$ ions at octahedral (B) sublattices due to their antiparallel alignment. The inclusion of $Ti^{4+}$ ions at the B-sites reduces the magnetic contribution from this sub-lattice, leading to a decrease in the total magnetization. Interestingly, despite the reduction in magnetization, the substitution of Ti has been shown to increase magnetostriction and strain sensitivity. This behaviour is attributed to the weakening of exchange interactions and modification of spin–orbit coupling associated with $Co^{2+}$ ions at octahedral sites. As a result, lower magnetic fields are required for domain wall motion, leading to higher strain sensitivity ($d\lambda/dH$). Such enhanced magnetostrictive response plays a crucial role in improving magnetoelectric coupling in composite systems. In Fig. 9 (a), it can be noted that the $M_r$, $M_s$, and $H_c$ values of all composite increase with increase in CTFO content, suggesting that the magnetic properties are strongly dependent on the content of CTFO. With increase in the sintering temperature $M_r$, and $M_s$ increase as can be clearly seen in Fig. 9 (c and d). Because enhance the sintering temperature reduced porosity and improvement of grains size which leads to a better magnetic alignment and stable magnetization. All magnetic parameters for various composition and sintering temperatures are listed in Table I.

**Table 1** The values of $P_s$, $P_r$, $E_c$, $M_s$, $M_r$, $H_c$ and $\alpha_{ME}$ for CTFO- BTO.

| **S. N** | **Composition** | **$P_s$** ($\mu C/cm^2$) | **$P_r$** ($\mu C/cm^2$) | **$E_c$** (kV/cm) | **$M_s$** (emu/g) | **$M_r$** (emu/g) | **$H_c$** (Oe) | **$\alpha_{ME}$** (mV/cm.Oe) |
|---|---|---|---|---|---|---|---|---|
| **A** | **1100°C** | | | | | | | |
| 1 | C10:B90 | 3.37 | 1.16 | 3.50 | 3.50 | 0.99 | 220.26 | 0.52 |
| 2 | C20:B80 | 2.78 | 0.76 | 2.80 | 8.66 | 2.59 | 261.07 | 0.59 |
| 3 | C30:B70 | 1.49 | 1.18 | 5.03 | 13.96 | 3.75 | 246.82 | 0.64 |
| **B** | **1200°C** | | | | | | | |
| 1 | C10:B90 | 7.78 | 1.59 | 1.82 | 4.09 | 1.19 | 255.39 | 0.77 |
| 2 | C20:B80 | 5.46 | 1.41 | 3.11 | 10.01 | 3.39 | 277.23 | 0.98 |
| 3 | C30:B70 | 2.38 | 1.05 | 4.41 | 16.65 | 4.78 | 265.29 | 1.28 |
| **C** | Pure Compound | | | | | | | |
| 1 | CTFO-(C) | | | | 67.54 | 6.56 | 87.86 | |
| 2 | BTO-(B) | 8.013 | 1.423 | 0.810 | | | | |

### 3.6 Magneto-electric coupling measurements

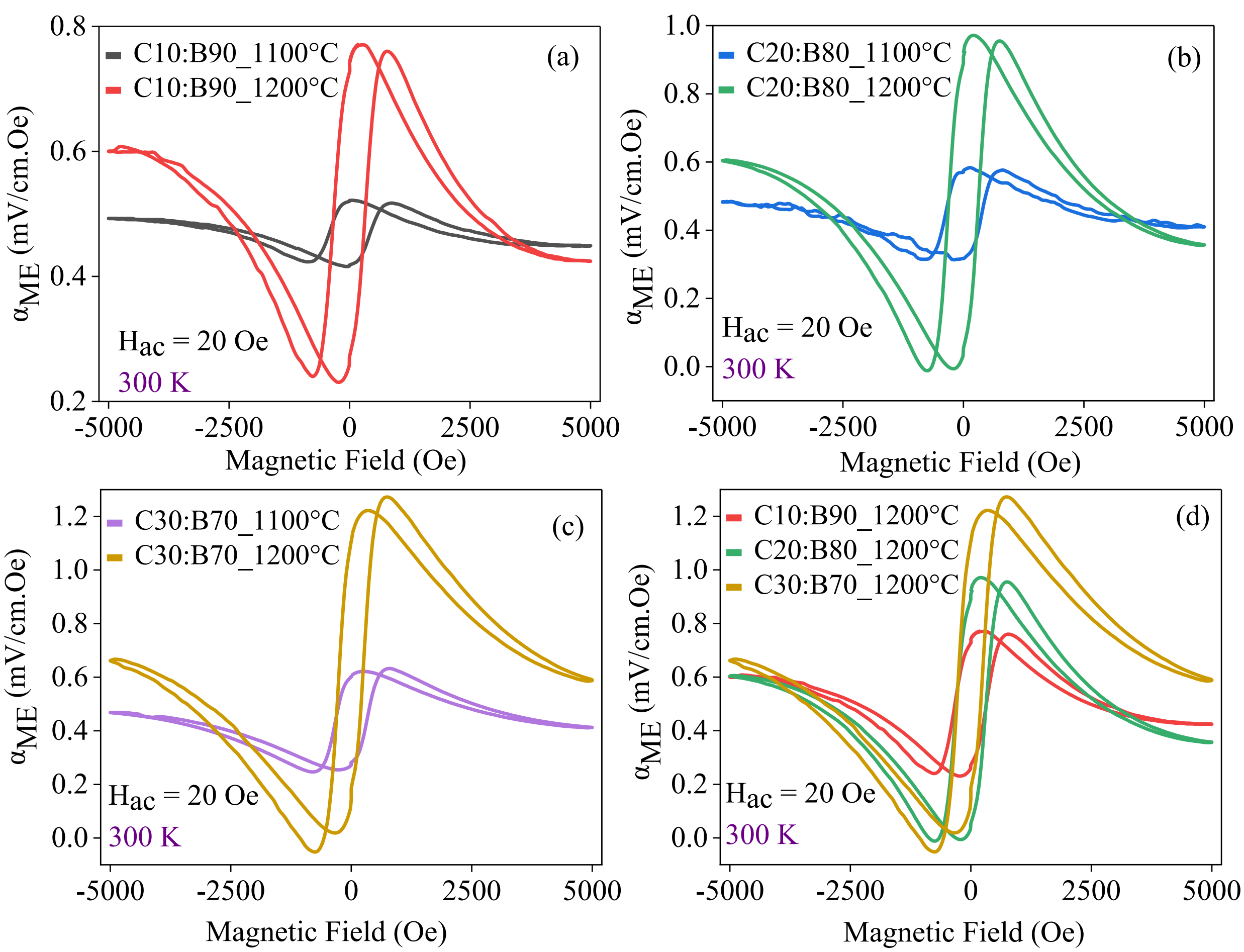

Fig. 10. Variation of magnetoelectric coefficient ($\alpha_{ME}$) with DC magnetic field for all composites. (a-c) show the effect of sintering temperature (1100 °C vs. 1200 °C) on ME coupling, while (d) compares the influence of CTFO concentration ($x$ = 10, 20, 30%) for the 1200 °C series.

The P-E and M-H loops study confirms the simultaneous presence of ferroelectric (FE) and ferromagnetic (FM) behaviours in the composite at room temperature. This indicates coupling between the FE and FM order parameters. The interaction occurs through magneto-mechanical-electric interactions and is known as the magneto-electric effect. Its strength is quantified by the ME voltage coefficient ($\alpha_{ME}$) [59]. It was measured using a dynamic lock-in technique and it was used to calculate the induced output voltage (dv) across the thickness (t) and the magnitude of the applied magnetic field (dH), and the ME coupling coefficient was calculated using the following equation,

$$\alpha_{ME} = \frac{dE}{dH} = \frac{dV}{dH.t} = \frac{V_{out}}{H_{ac}.t} \quad \text{(iii)}$$

Here $V_{out}$ is the measured voltage across the sample, $H_{ac}$ is the applied AC magnetic field, t is the thickness of composite [12,43,60]. The ME coefficient arises from the strain-mediated coupling between the magnetostrictive (CTFO) and piezoelectric (BTO) phases. When a magnetic field is applied, magnetostrictive deformation occurs in the CTFO phase, which is elastically transferred to the BTO phase, generating electric polarization through the piezoelectric effect. The ME coefficient can be expressed as:

$$\alpha_{ME} \propto qdk \qquad \text{(iv)}$$

Where d is the piezoelectric coefficient of BTO phase, q is the piezomagnetic coefficient of CTFO phase: q = dλ/dH (λ is the magnetostriction) and $k$ is the interfacial coupling efficiency between both phases of the ME material [2,12,61]. Fig. 10 (a-c) illustrates the magneto-electric (ME) coupling coefficient ($\alpha_{ME}$) of all composites as a function of the applied DC magnetic field ($H_{dc}$), varying between ± 5 kOe, at a fixed frequency of 23 Hz and applied constant ac magnetic field of 20 Oe. All composites exhibit a typical asymmetric behaviour in ME loop, this phenomenon is directly attributed to the magnetic remanence and coercive field of the ferromagnetic CTFO phase. Maximum ME coupling in both cycle (positive and negative magnetic field) indicates that strain mediated coupling.

We observed that all composites show coupling between both orders (FE & FM), and the coupling coefficient are listed in Table 1. The $\alpha_{ME}$ coupling of all composites initially rises to a maximum and then decreases, with the dc magnetic field. The increasing behaviour can be attributed to the magnetostriction coefficient, which increases with the applied magnetic field and reaches saturation at a certain field value causing $\alpha_{ME}$ to decline, eventually reaching very low values at higher dc magnetic fields. ME coupling is affected by composition and sintering temperature. With increasing ferromagnetic content enhancing the magnetostrictive contribution in the composite, while reducing the ferroelectric contribution. CTFO content increases electrical conductivity, leading to higher leakage current and reduced polarization. Therefore, an optimal balance between these two phases helps to achieve maximum ME coupling. Through microstructural and interfacial engineering, the composite Ti-substituted $CoFe_2O_4$–$BaTiO_3$ exhibits superior performance at low frequencies (~23 Hz). Fig. 10 (c) shows that composite C30:B70 (x = 30) exhibits the highest ME coefficient. At low CTFO content (x = 10), the magnetostrictive contribution is insufficient, while at intermediate composition (x = 20), both phases contribute moderately but do not achieve optimal coupling. At x = 30, there is a sufficient contribution to generate magnetostrictive strain, resulting in maximum ME coupling. The ME effect in the composite arises from the strain-mediated coupling between the ferroelectric (BTO) and ferromagnetic (CTFO) phases. When a magnetic field is applied, the CTFO phase undergoes magnetostrictive strain, which is elastically transferred to the BTO phase, where it creates electric polarization through the piezoelectric effect. Fig. 10 (d) shows the C30:B70_1200 °C composite which shows the most efficient ME response (~1.28 mV/cm.Oe). When the sintering temperature is raised from 1100 °C to 1200 °C, $\alpha_{ME}$ increased. This is related to reduced porosity and grain growth, leading to larger grain sizes and reduced grain boundaries, resulting in more effective elastic coupling and lower mechanical losses at the CTFO-BTO interface. It is observed that maximum ME coupling does not correspond to the highest polarization or magnetization individually. Instead, $\alpha_{ME}$ is maximized when the product of piezoelectric coefficient (d) and piezomagnetic coefficient (q) and interfacial coupling efficiency (k) is optimized. Although composite C30:B70_1200°C does not show the highest polarization, it still has the highest ME coefficient due to its enhanced magnetic response and better interfacial coupling. These results clearly indicate that a high concentration of CTFO (30%) provides an optimal balance for ensuring the transfer of magnetostrictive strain to the piezoelectric BTO phase to produce maximum magneto-electric coupling in this series, demonstrating the essential role of microstructure and interface engineering in optimizing magneto-electric coupling.

**Table. 2** - Comparison of the value of $\alpha_{ME}$ as obtained in the present work with that of some reported literature concerning particulate ME composites

| System (Bulk Composite) | $H_{DC}$ (Oe) | Freq (Hz) | $\alpha_{ME}$ (mV /cm.Oe) | Reference |
|---|---|---|---|---|
| 0.4 $CoFe_2O_4$–0.6 $BaTiO_3$ | 2500 | 500 | 0.42 | [19] |
| 0.9PVDF-0.1 $CoFe_2O_4$ | 1500 | 540 | 13.20 | [62] |
| 0.5 $BaTiO_3$–0.5 $CoFe_2O_4$ | 1800 | | 1.4 | [63] |
| $BaTiO_3$– $CoFe_2O_4$ | 1000 | 1000 | 0.08 | [64] |
| 0.9$BaTiO_3$–0.1$CoFe_2O_4$ | 2000 | 273 | 4.2 | [54] |
| Ni–$BaTiO_3$ | 2500 | 850 | 18.3 | [65] |
| 0.8 $BaTiO_3$–0.2 $CoFe_2O_4$ | — | — | 0.98 | [18] |
| 0.94 $Ba_{0.99}Tb_{0.02}Ti_{0.99}O_3$ –0.06 $Co_{0.94}Tb_{0.06}Fe_2O_4$ | 580 | 1000 | 2.81 | [66] |
| $BaTi_{0.89}Sn_{0.11}O_3$–$CoFe_{1.9}Bi_{0.1}O_3$ | 1420 | 1000 | 7.8 | [67] |
| $Ba_{0.99}Dy_{0.02}Ti_{0.99}O_3$–$CoFe_{1.9}Zn_{0.1}O_4$ | 1590 | 1000 | 1.07 | [68] |
| 30$Co_{1.2}Ti_{0.2}Fe_{1.6}O_4$–70$BaTiO_3$ | 1040 | 23 | **1.28** | This work |

## 4 Conclusions

We successfully synthesized lead-free (x)$Co_{1.2}Ti_{0.2}Fe_{1.6}O_4$-(100-x)$BaTiO_3$ multiferroic composites with x = 10, 20, and 30, sintered at 1100 and 1200 °C. X-ray diffraction analysis confirmed the coexistence of CTFO and BTO phases. Microstructural analysis showed that the grains were well densified, at a higher sintering temperature of 1200 °C. Magnetic (M-H) and ferroelectric (P-E) measurements confirmed the coexistence of ferromagnetic and ferroelectric phases, thereby confirming the multiferroic nature of the synthesized composite. Composites sintered at higher temperatures exhibited a higher dielectric constant, lower dielectric loss, and increased magnetization and polarization. The C30:B70 composite sintered at 1200 °C exhibited the highest ME coefficient (~1.28 mV/cm.Oe), attributed to the optimal balance between magnetostrictive and piezoelectric contributions and enhanced interfacial coupling. The results highlight that strong magneto-electric coupling is governed by coupling efficiency rather than individual phase properties. Overall, present results indicate that microstructural engineering through thermal treatment is an important tool for maximizing multifunctional properties in lead-free multiferroic systems. The results provide valuable information for designing high-performance lead-free magnetoelectric materials for sensor and energy harvesting applications.

**Acknowledgments**

S.G. acknowledges the financial support from the Anusandhan National Research Foundation (ANRF), New Delhi, under project SUR/2022/004713.